\documentclass[conference]{IEEEtran}

\usepackage{amsmath,amsfonts}
\usepackage{algorithmic}
\usepackage{algorithm}
\usepackage{subfigure}

\usepackage{array}
\usepackage[caption=false,font=normalsize,labelfont=sf,textfont=sf]{subfig}
\usepackage{textcomp}
\usepackage{stfloats}
\usepackage{url}
\usepackage{verbatim}
\usepackage{graphicx}
\usepackage{cite}
\usepackage{color}
\begin{document}

\title{\huge Computation Offloading for Multi-server Multi-access Edge Vehicular Networks: A DDQN-based Method}

\vspace{-8mm}

\author{
\IEEEauthorblockN{Siyu Wang\IEEEauthorrefmark{1}, Bo Yang\IEEEauthorrefmark{1}, Zhiwen Yu\IEEEauthorrefmark{1},  Xuelin Cao\IEEEauthorrefmark{2}, Yan Zhang\IEEEauthorrefmark{3},
 and Chau Yuen\IEEEauthorrefmark{4}} 
  
\IEEEauthorblockA{\IEEEauthorrefmark{1}School of Computer Science, Northwestern Polytechnical University, Xi'an, Shaanxi, 710129, China} 
\IEEEauthorblockA{\IEEEauthorrefmark{2}School of Cyber Engineering, Xidian University, Xi'an, Shaanxi, 710071, China}
\IEEEauthorblockA{\IEEEauthorrefmark{3}Department of Informatics, University of Oslo, 0316
Oslo, Norway}
\IEEEauthorblockA{\IEEEauthorrefmark{4}School of Electrical and Electronics Engineering, Nanyang Technological University, Singapore}
}

\maketitle

\begin{abstract}
In this paper, we investigate a multi-user offloading problem in the overlapping domain of a multi-server mobile edge computing system. We divide the original problem into two stages: the offloading decision making stage and the request scheduling stage. To prevent the terminal from going out of service area during offloading, we consider the mobility parameter of the terminal according to the human behaviour model when making the offloading decision, and then introduce a server evaluation mechanism based on both the mobility parameter and the server load to select the optimal offloading server. In order to fully utilise the server resources, we design a double deep Q-network (DDQN)-based reward evaluation algorithm that considers the priority of tasks when scheduling offload requests. Finally, numerical simulations are conducted to verify that our proposed method outperforms traditional mathematical computation methods as well as the DQN algorithm.
\end{abstract}

\begin{IEEEkeywords}
Computation offloading, multi-access edge computing, vehicular networks, double deep Q-network\footnote{This paper will be presented at VTC-Spring 2024, Singapore.}
\end{IEEEkeywords}

\section{Introduction}
\begin{figure}[t]
\centerline{\includegraphics[width=0.95\linewidth]{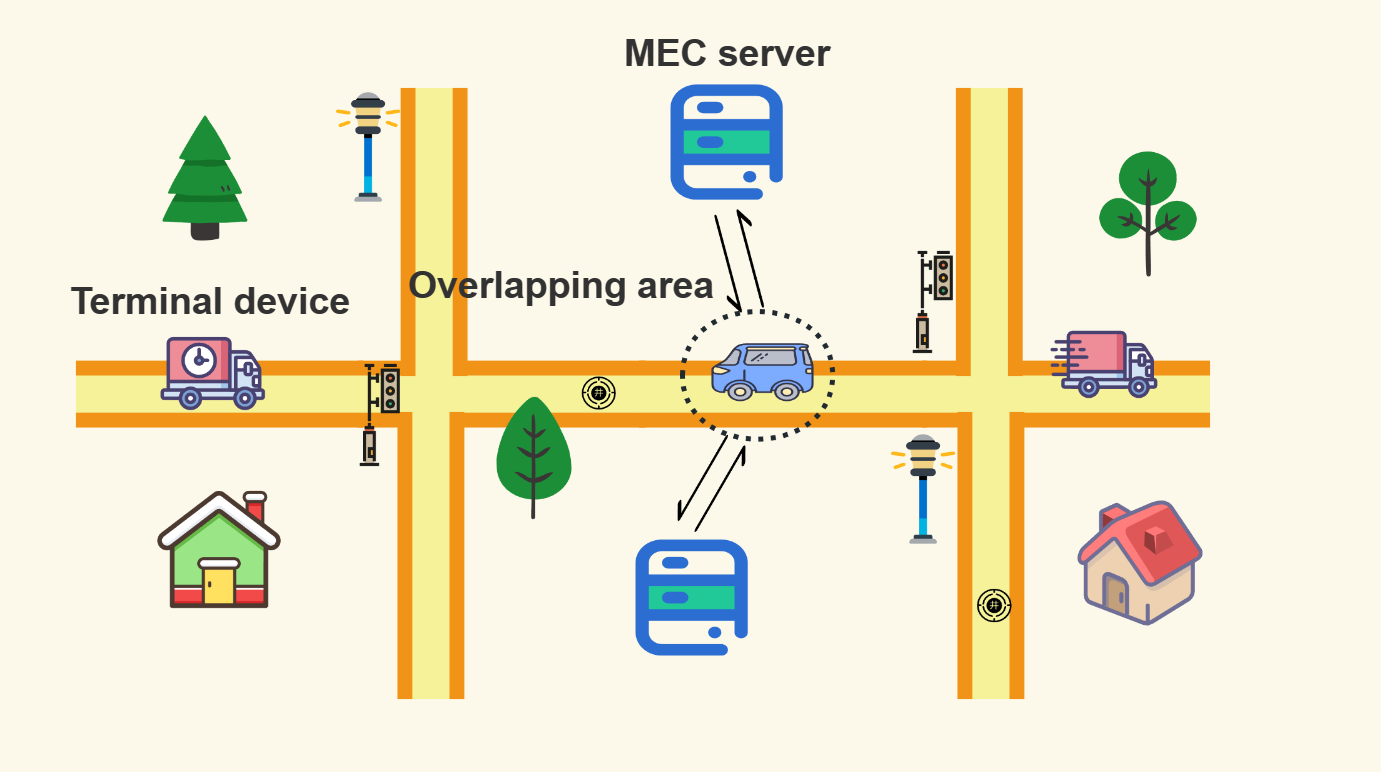}}
\caption{The considered multi-server multi-access edge vehicular network scenario, where the terminal device (e.g., the blue vehicle) within the overlapped area needs to choose the optimal offloading strategy.}
\label{system}
\end{figure}

\IEEEPARstart{W}{ith} the development of Multi-access Edge Computing (MEC) technology, MEC servers are moving closer to the terminal devices (TDs), which can be served more efficiently as the transmission latency is greatly reduced~\cite{b1}. In this case, it becomes possible to run the latency sensitive and computationally intensive applications on TDs with limited computing and storage resources. However, in multi-access edge vehicular networks, the mobility of the TDs will greatly affect their quality of service (QoS) during task offloading. In particular, frequent movement of TDs may cause some of them to be out of range of the MEC server, as illustrated in Fig.~\ref{system}, thereby resulting in service interruption. Meanwhile, individual tasks have different characteristics, such as task size and computing resources required. In this context, the mobility of the TDs and the characteristics of the tasks should be considered together when optimising the offloading decisions and resource allocation. 

\textbf{Related work.} In order to improve the service experience for users, some existing work has been proposed based on different optimization objectives. Specifically, the authors of \cite{b2} proposed a multi-task learning-based feed-forward neural network (MTFNN) model that was trained to jointly optimize the offload decision and computational resource allocation. The authors of \cite{b3} proposed an effective task scheduling algorithm based on dynamic priority, which significantly reduced task completion time and improved QoS. In \cite{b4}, the authors proposed a hybrid task offloading scheme based on deep reinforcement learning that achieved vehicle-to-edge and vehicle-to-vehicle offloading by jointly considering delay constraints and resource demand. In \cite{b5}, an efficient task offloading scheme for cellular vehicle-to-everything was proposed to improve the offloading reliability and latency. %The authors of \cite{b30} developed a cooperative three-layer vehicle-assisted multi-access edge computing network to improve the robustness of computational offloading strategies in a high mobility environment.  
In addition, the offloading ratio and bit allocation of multiple vehicles were jointly optimised to adapt to the dynamic environment~\cite{b6}. The authors of \cite{b7} studied the mobility-aware computation offloading design for MEC-based vehicular wireless networks, where random mobility of vehicles is considered. To improve the offloading efficiency, a Q-learning based approach was proposed to investigate the trade-off between delay demand and energy consumption during the task offloading process~\cite{b8}. Notably, few of the above works jointly consider the mobility and task features (especially task priority) when making offloading decisions and resource allocation, thereby reducing the offloading efficiency.

\textbf{Contributions.} In this paper, we present an offloading scheme for multi-server multi-access edge vehicular networks by jointly considering the mobility and task priority of TDs. In order to fully exploit the server resources, a double deep Q-network (DDQN)-based reward evaluation algorithm is further proposed to handle a large number of offloading requests when the number of devices gradually increases. Numerical results show that our proposed scheme outperforms the traditional mathematical-based methods as well as the DQN-based algorithm in terms of the number of important tasks accomplished by the system per unit of time.

\section{System Model}
\subsection{Overview}
The multi-access edge vehicular system considered in this paper is shown in Fig.~\ref{system}. The MEC system includes some terminal devices (TDs) that send offloading requests to the MEC servers. Let $M$ be the total number of TDs and $N$ be the number of MEC servers, ${\cal U}=\left \{ U_1, U_2,...,U_M \right \}$ and ${\cal E}=\left \{ E_1, E_2,...,E_N \right \}$ denote the set of TDs and MEC servers respectively.  Each TD can move within the overlapping area of multiple MEC servers and therefore needs to dynamically select the optimal offloading strategy.

Consider a given time period containing a total of $T$ time slots, defined as $t_1, t_2, ..., t_T$. During the considered time period, we define the set of tasks generated by the total $M$ terminals as ${\cal J} = \left \{\textbf{J}_1, . .., \textbf{J}_m, ..., \textbf{J}_M \right \}$, where $\textbf{J}_m = \{J_m(1), J_m(2), ..., J_m(T)\}, \ \forall m \in [1,M]$. Then the corresponding two-dimensional offloading decision set is denoted as ${\cal D}= \left \{\textbf{D}_1, ... , \textbf{D}_m, ..., \textbf{D}_M \right \}$, where $\textbf{D}_m = \{D_m(1), D_m(2), ..., D_m(T)\}, \ \forall m \in [1,M]$. The set of selected optimal MEC servers is ${\cal K} =\left \{\textbf{K}_1, ..., \textbf{K}_m, ..., \textbf{K}_M \right \}$, where $\textbf{K}_m = \{K_m(1), K_m(2), ..., K_m(T)\}, \ \forall m \in [1,M]$. For $U_m$ at a given time $t$, if the task is offloaded to the server, then we have $D_m(t)=1$ and the optimal MEC server is $K_m(t)$, otherwise $D_m(t)=0$.

%\textcolor{blue}{two integer variables $loc\underline{~}x\in \left ( 0,40 \right ) $ and $loc\underline{~}y\in \left ( 0,40 \right )$ represent the horizontal and vertical coordinates of a terminal device or MEC server. $M\underline{~}r$ indicates the movement rate of the terminal device. $M\underline{~}d\in \left ( 0,1,2,3 \right ) $ indicates the movement direction of the terminal device. The variable $M\underline{~}d$ is randomly initialized and has a certain probability of not changing according to the human orientation preferences. The service radius of a MEC server is represented by the symbol $R$. The set $T=\left \{ 1,2,...,T \right \} $ marks the time slot \cite{b5}.}

\subsection{Terminal Device Model}
\begin{figure}[thb]
\centerline{\includegraphics[width=0.98\linewidth]{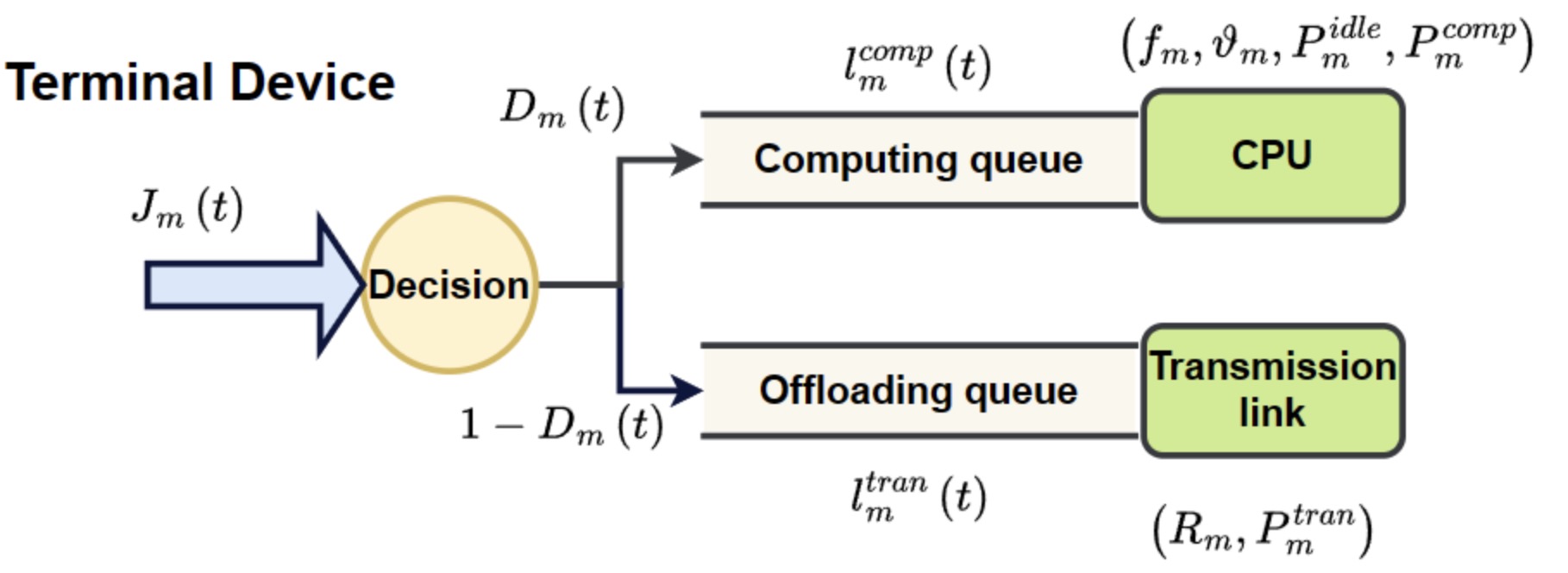}}
\caption{Terminal device model.}
\label{terminal}
\end{figure}

At a given time slot $t$, the $m$-th TD (denoted as $U_m$) generates an indivisible task $J_m(t)$. To illustrate the features of the task, $z\left(J_m(t)\right)$ (in bits) is used to represent the size of $J_m(t)$. As shown in Fig.~\ref{terminal}, each TD has two queues, one is the computing queue and the other is the offloading queue\cite{b9}. We let a binary variable $D_{m}(t) \in \left \{ 0,1 \right \} $ represent the result of the offloading decision. That is to say, if the revenue evaluation is positive, $D_{m}(t)= 0$ holds and then the task $J_m(t)$ can be added to the offloading queue, otherwise, the task $J_m(t)$ will be sent to the computing queue of the selected MEC server.

For the $m$-th terminal $U_m$ at the time slot $t$, the length of the computing queue is expressed by $l^{comp}_m(t)$, which reflects the current load of the queue in bytes. $f_{m}$ indicates the CPU cycle frequency (i.e., CPU cycles per second) of $U_m$. $\vartheta_{m}$ indicates the maximum number of data bits that can be processed simultaneously by each CPU clock cycle (i.e., bits per CPU cycle). $P^{comp}_{m}$ and $P^{idle}_{m}$ represent the power consumption of $U_m$ that stays at the computing state and idle state, respectively. 

\textbf{Computing queue.} Suppose that the task is processed by $U_m$ itself, then the local processing delay includes the queuing delay and task processing delay and can be calculated as
$\tau_m^{L}(t)\!=\! \frac{\left ( l^{comp}_{m}(t)+z\left(J_m(t)\right)\right )}{f_{m} \vartheta_{m}}$. Similarly, the energy consumption of local processing is calculated as $\varepsilon_m^L(t)= P^{comp}_{m}\tau_m^{L}(t)$, and the energy consumption for idle can be obtained as 
$\varepsilon_m^I(t)=P^{idle}_{m} \tau_m^{K_m(t)}$, where ${K_m}\in {\cal S}$ indicates the selected MEC server of $U_m$, $\tau_m^{K_m(t)}$ indicates the delay raised by the task processing at the selected MEC server of $U_m$, which will be introduced in the following part.

\textbf{Offloading queue.} The length of the offloading queue is expressed by $l^{tran}_{m}(t)$ and $P^{tran}_{m}$ represents the transmission power of $U_m$. Then the achieved data rate is calculated as $R_{m}=W \log_{2}{\left ( 1+P^{tran}_{m} \left | h_{m,s}  \right |^{2}/{\sigma ^{2} }\right ) }$, where $W$ indicates the bandwidth of the offloading link, $\left | h_{m,s}  \right |^{2}$ indicates the channel gain between the terminal device and the MEC server, $\sigma ^{2}$ indicates the noise power. 

In this case, the transmission delay is calculated as
$\tau_m^{tran}(t)=\left (l^{tran}_{m}(t)+z\left(J_m(t)\right) \right)/R_{m}$,
and the corresponding energy consumption is obtained as
\begin{equation} \label{2}
    \varepsilon_m^{tran}(t)\!=\!P^{tran}_{m} \tau_m^{tran}(t)\!=\!\frac{P^{tran}_{m}\left (l^{tran}_{m}(t)\!+\!z\left(J_m(t)\right) \right)}{R_{m}}.
\end{equation}

Therefore, the total energy consumption of $U_m$ during the offloading is calculated by
\begin{equation} \label{3}
   \varepsilon_m^{off}(t)=\varepsilon_m^{tran}(t)+\varepsilon_m^I(t).
\end{equation}

By substituting $\varepsilon_m^I(t)$, $\varepsilon_m^{tran}(t)$ and $R_{m}$ into (\ref{3}), we have
\begin{equation} \label{4}
   \varepsilon_m^{off}(t)\!=\!\frac{P^{tran}_{m}\left (l^{tran}_{m}(t)+z\left(J_m(t)\right) \right)}{W \log_{2}{\left ( 1\!+\!P^{tran}_{m} \left | h_{m,s}  \right |^{2}/{\sigma ^{2} }\right ) }}+P^{idle}_{m} \tau_m^{K_m(t)}.
\end{equation}

\subsection{MEC Server Model}
%At a given time slot $t$, the MEC server will broadcast the current load of the task queue to all the terminal devices, 
\begin{figure}[thb]
\centerline{\includegraphics[width=0.8\linewidth]{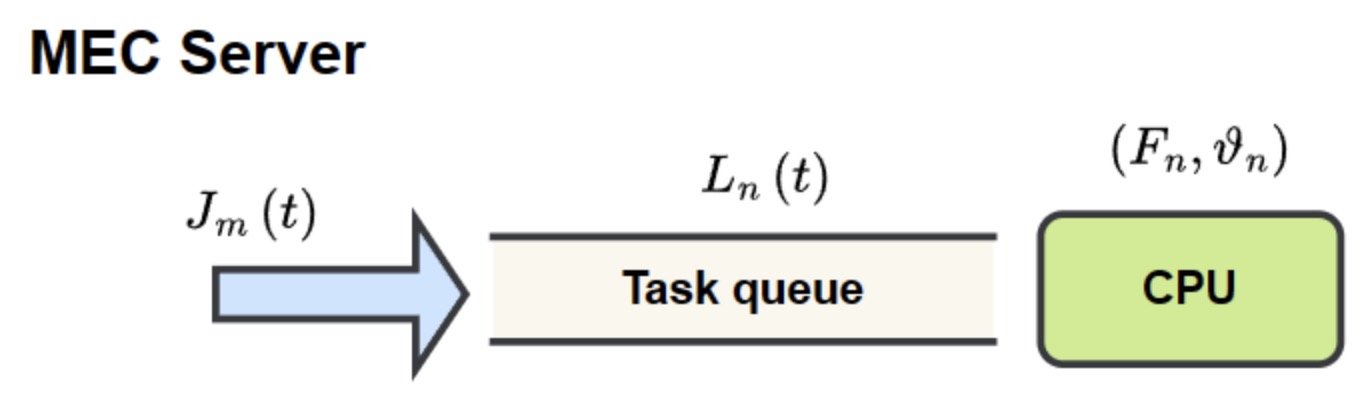}}
\caption{MEC server model.}
\label{server}
\end{figure}

As shown in Fig.~\ref{server}, the MEC server has a task queue and all tasks are processed according to the first-come first-served rule. Specifically, the length of the task queue is given by $L_n(t)$. $F_{n}$ represents the CPU frequency of the MEC server $E_n, \forall n\in [1, N]$. $\vartheta_{n}$ represents the maximum number of data bits that can be processed simultaneously by the MEC server $E_n$. The delay of the task processing on the server is therefore equal to $\tau_m^{K_m(t)}$, which is calculated as
$\tau_m^{K_m(t)}=\left ( L_{n}(t) + z(J_m(t)\right )/\left ( F_{n} \vartheta_{n}  \right )$.

Considering that the size of the result returned by the MEC server after task processing is usually very small, the downloading delay can be ignored in this paper. Therefore, the total delay of the offloading procedure is calculated as %\vspace{-2mm}
\begin{equation} 
 \begin{aligned}
 \tau_{m}^{off}(t)&=\tau_m^{tran}(t)+\tau_m^{K_m(t)}\\
 &=\frac{\left (l^{tran}_{m}(t)\!+\!z\left(J_m(t)\right) \right)}{W \log_{2}{\left ( 1\!+\!P^{tran}_{m} \left | h_{m,s}  \right |^{2}/{\sigma ^{2} }\right ) }}\!+\!\frac{L_{n}(t) \!+\! z\left(J_m(t)\right)}{F_{n} \vartheta_{n}}.
\end{aligned}
\end{equation}

\section{Problem Formulation and Analysis}
\subsection{Problem formulation}
%The goal of the system is to accomplish as many important tasks as possible in a period. 

%For ease of understanding, we assume that the sets of tasks processed locally by the devices are $J_{1}^{l}  =\left \{ J_{1}\left ( x \right ), J_{1}\left ( y \right ),..., J_{1}\left ( z \right ) \right \},..., J_{M}^{l}  =\left \{ J_{M}\left ( x \right ), J_{M}\left ( y \right ),..., J_{M}\left ( z \right ) \right \}$ respectively. We also assume that the sets of tasks processed by the servers are $J_{1}^{s}  =\left \{ J_{1}\left ( a \right ), J_{1}\left ( b \right ),..., J_{1}\left ( c \right ) \right \},..., J_{S}^{s}  =\left \{ J_{S}\left ( a \right ), J_{S}\left ( b \right ),..., J_{S}\left ( c \right ) \right \}$ respectively. Because the resources of the server are limited, it cannot handle all the offloading tasks, the server generates a scheduling policy for processing as many important tasks as possible. These scheduling policies are defined as $D^{s}=\left \{D_{1}^{s} \left ( a \right ),...,D_{1}^{s} \left ( b \right ),..., D_{S}^{s} \left ( a \right ),...,D_{S}^{s} \left ( b \right )\right \}$.
Given the considered time period with total $T$ time slots, the number of tasks that $U_{m}$ process locally at time slot $t$ is defined as $\xi_{m}\left ( t \right )$, which should satisfy the following condition
\begin{equation}
    \sum_{i=1}^{\xi_{m}\left ( t \right )} z\left ({q^{comp}_m(t)[i]}\right )\le \pi \vartheta _{m} f_{m}, 
\end{equation}
where $\pi$ indicates the time slot size, $q^{comp}_m(t)[i]$ denotes the $i$-th element of the computing queue of terminal device $U_{m}$, $z(q^{comp}_m(t)[i])$ denotes the size of the element.

Similarly, the number of tasks that $E_{n}$ process at time slot $t$ is defined as $\xi_{n}\left ( t \right )$, which should satisfy the following condition
\begin{equation}
    \sum_{i=1}^{\xi_{n}\left ( t \right )} z\left ({q_n(t)[i]}\right )\le \pi  \vartheta _{n}  F_{n}, \label{eq6}
\end{equation}
where ${q_n(t)[i]}$ denotes the $i$-th element of the task queue of MEC server $n$. 

Without loss of generality, we consider that different tasks have different priorities. The priority of the task $J_{m}\left ( t \right )$ is denoted as $p\left(J_{m}\left ( t \right )\right)$. Therefore, to illustrate the importance coefficient, we introduce the variable $I$ to represent the criterion of handling important tasks as below
%\vspace{-2mm}
\begin{equation}
 \begin{aligned}
   I= \frac{1}{T} \sum_{t=1}^{T} \left({\sum_{m=1}^{M}\sum_{i=1}^{\xi_{m}\left ( t \right )  } p\left ({q^{comp}_m(t)[i]}\right ) }\right. \\
  \;\;\;\;\;\;\;\;\;\;\;\; \left. { + \sum_{n=1}^{N}\sum_{i=1}^{\xi_{n}\left ( t \right )}p\left ({{q_n(t)[i]}}\right )}\right ).\\
    \end{aligned}
\end{equation}

Our object is to find the optimal combination of $\cal D$ and $\cal K$ to maximize $I$ under some constraints. The optimization problem can be formulated as 
\begin{subequations} 
\begin{align}
& \;\;\;\; \mathbb{P}: \ \underset{{\cal D, \cal K}}{\rm max}\;\;  I  \notag \\
& \;\;\;\;\;\;{{s}}{{.t}}{\rm{.}}\;\;\; \; D_{m}\left ( t\right )  \in \{0,1\}, \ 1 \leq m \leq M,\\
&\;\;\;\;\;\;\;\;\; \;\;\;\;\;\; K_{m}\left ( t\right ) \in {\cal E}, \ 1 \leq m \leq M.
 \end{align}
\end{subequations}

% \begin{subequations}\label{eq:ctr_shale}
% \begin{align}
% arg\max_{\pi_{m} }I   \ \ \tag{\ref{eq:ctr_shale}}
% \end{align} 
% \begin{alignat}{2}
% \text{s.t.} &d\in \left \{ 0,1 \right \} , \space d^{s} \in \left \{ 0,1 \right \} ,\space \forall d\in D ,\space \forall d^{s} \in D^{s} \\
% &k\in \left [ 1,S \right ]  ,\space \forall k \in Best\underline{~}server\\
% & \sum_{i=0}^{N_{x}^{l}-1 } J_{x}^{l}\left [ i \right ] .size\le F_{x}^{comp} \times P_{m}\times t     \\
%     &\sum_{i=0}^{N_{y}^{s}-1 } J_{y}^{s}\left [ i \right ].size\le F_{y}^{edge} \times P_{s}\times t  
% \end{alignat}
% \end{subequations}

In this paper, the resource allocation problem can be divided into two stages. The first stage is the offloading decision stage, where each terminal device makes two-dimensional offloading decisions according to the calculated benefit and then selects the optimal offloading server and send offloading requests. The second stage is performed by the MEC server to run a reward evaluation algorithm based on DDQN to generate appropriate scheduling results.

\subsection{Problem Analysis}
\subsubsection{Offloading decision making} 
When a new task is generated, the terminal device decides whether to process the task locally or offload it to a nearby edge server. The terminal device calculates the delay saving and energy saving respectively\cite{b10}. We define the delay benefit as $\tau_m^{+  }(t)=\tau_m^{L}(t)-\tau_{m}^{off}(t)$, and the energy benefit is defined as $\varepsilon_m^{+}(t)=\varepsilon_m^L(t)- \varepsilon_m^{off}(t)$. If the total benefit is positive, the task can be offloaded, otherwise, the task will be processed locally.

We also introduce a variable $\lambda \!\in\! \left [ 0, 1 \right ]$, to represent the preference factor that determines whether the optimization objective focuses more on delay reduction or energy reduction. We let $o _m^{+ }(t)$ denote the overall benefit evaluation result, which is given by \vspace{-1mm}
\begin{equation}
    o _m^{+  }(t) = \lambda \tau_m^{+  }(t)+\left ( 1-  \lambda   \right ) \varepsilon_m^{+}(t)\label{eq}
\end{equation}

Therefore, the variable $D_{m}\left ( t \right )  \in \left \{ 0,1 \right \} $ that represents the result of the offloading decision can be determined by 
\begin{equation}
    D_{m}\left ( t \right ) =\begin{array}{l}  \left\{\begin{matrix}   0, \ o _m^{+  }(t)> 0 \\   1, \ o _m^{+  }(t)\le  0 \end{matrix}\right.    \end{array}.  \label{eq}
\end{equation}

Having made a two-dimensional decision, we move on to the next decision-making process to select an optimal server to send the offloading requests to. We let $\left ( x_{m}(t), y_{m}(t) \right )$, $\left ( x_{m}(t+1), y_{m} (t+1) \right )$ denote the position of the terminal $U_{m}$ at time $t$ and $t+1$ respectively. $x_{n}$ and $y_{n}$ give the horizontal and vertical coordinates of the MEC server $E_{n}$.
$d_{m,n}\left ( t \right )$ and $d_{m,n}\left ( t+1 \right )$ represent the distance between the terminal device $U_{m}$ and the MEC server $E_{n}$ at the time slot $t$ and $t+1$, respectively.
The process of selecting the optimal offloading server involves three steps. Firstly, an MEC server is selected to calculate the distance between the terminal and the MEC server at time slot $t$. Second, calculate the distance between the terminal and the selected MEC server at the next time slot $t+1$. Thirdly, calculate the offloading score of the server, i.e., $score_{m,n}(t)=\left ( d_{m,n}\left ( t \right )- d_{m,n}\left ( t+1 \right ) \right )+\alpha o _m^{+ }(t)$, where $\alpha$ is the scaling factor that is positively related to the movement rate of the terminal. The optimal MEC server with the highest score is selected.

\subsubsection{Offloading request scheduling}
\begin{figure}[t]
\centerline{\includegraphics[width=1.01\linewidth]{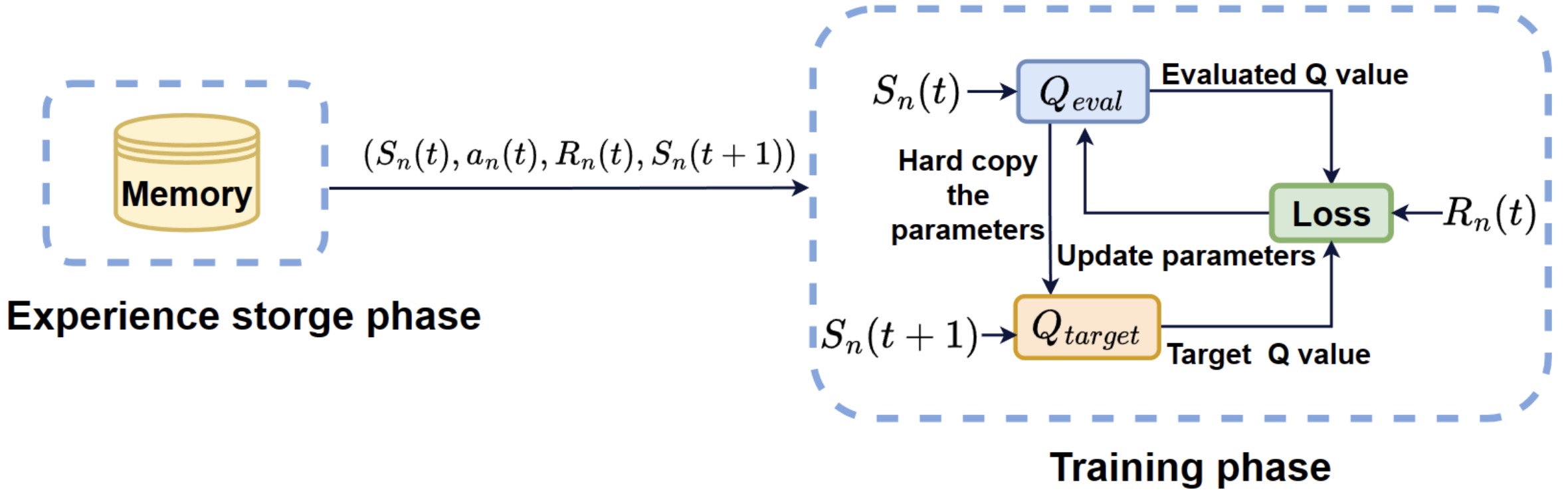}}
\caption{The proposed R-DDQN algorithm flow.}
\label{algorithm}
\end{figure}

As the offloading decision is generated distributively by each terminal, it may result in multiple terminals selecting the same optimal offload server at the same time. This leads to increased transmission delay due to channel congestion and inaccurate revenue evaluation. To solve this problem, we propose a DDQN-based offloading requests scheduling algorithm that is implemented at each MEC server to schedule multiple offloading requests and thus fully utilize the limited resources of the server to handle more important tasks. Firstly, the MEC server $E_{n}$ parses the received offloading requests to obtain the information about the offloading task, such as the task priority, the required computing resources and so on. Secondly, the server $E_{n}$ evaluates its remaining computing resources at the time slot $t$, which is given by $\rho_{n}(t)\!=\! 1 - \frac{L_n(t)}{F_{n} \pi \vartheta_{n}}$. Thirdly, we introduce a reward evaluation algorithm to obtain the optimal scheduling result, as shown in Fig.~\ref{algorithm}, and then the scheduling result will be sent to each terminal~\cite{b11}.

We let $\phi_{n}\left ( t \right )$ denote the set of the received offloading tasks at the MEC server $E_{n}$ within the time slot $t$, then the state space and the action space can be given by
\begin{equation}
    S{n}\left ( t \right ) =\left \{ z(\phi_{n}\left ( t \right )), p(\phi_{n}\left ( t \right )),\rho_{n}\left ( t \right )\right \}, \label{eq}
\end{equation}
\begin{equation}
    A_{n}\left ( t \right ) =\left \{ 0,1 \right \}^{M}.  \label{eq}
\end{equation}

If the terminal has not sent any offload requests or the MEC server has rejected the terminal's offloading request, the value of the element whose subscript corresponds to the terminal's serial number is 0, otherwise its value is 1. Taking the task $J_{m}\left ( t \right )  $ as an example, the variable $v_{b}$ is set as the base reward value, and the reward value obtained by completing $J_{m}\left ( t \right )$ is defined as \vspace{-0.5mm}
\begin{equation}
    r_{m}\left ( t \right )=v_{b} p_{m}\left ( t \right ).  \label{eq}
\end{equation}

We define $C_{r_{n} }\left ( t \right )$ as the immediate benefit of the MEC server $E_{n}$. According to \eqref{eq6}, the number of tasks that MEC server $E_{n}$ can process at the time slot $t$ (i.e., $\xi_{n}\left ( t \right )$) can be obtained, so we have  \vspace{-1mm}
\begin{equation}
    C_{r_{n} }\left ( t \right )=\sum_{i=1}^{\xi_{n}\left ( t \right )}p\left ({{q_n(t)[i]}}\right ) v_{b}.\label{eq}
\end{equation}

 Let ${\cal N}$ be the number of slots in which the task is waiting to be processed, $m$ is the number of offloading requests, so the expected benefit is defined as 
\begin{equation}
    E_{r_{n}} \left ( t \right ) =\sum_{m= 1}^{M} r_{m}\left ( t \right ) \frac{A_{n}\left ( t \right ) \left [ m \right ]}{\cal N}. \label{eq}
\end{equation}

The reward function is defined as $R_{n}(t)=C_{r_{n} } (t)+ E_{r_{n}} (t)$, which considers both the effect of the decision on the present and the future payoff of the offloading task.

The flow of the reward evaluation algorithm based on DDQN is shown in Fig.~\ref{algorithm}, the algorithm has two phases, including the experience storage phase and the learning phase \cite{b12}. In the experience storage phase, at time $t$, the MEC server $E_{n}$ first obtains the environment state $S_{n}(t) $. The $Q_{eval}$ network outputs the value of $Q_{eval}\left ( S_{n}(t), A_{n}\left ( t \right ) \right ) $ for all possible values of $A_{n}\left ( t \right )$. The agent uses the $\epsilon$-greedy strategy according to the results of the $Q_{eval}$ network, chooses an action $a_{n}(t)$ and receives the instant rewards $R_{n}(t)$. At time slot $t\!+\!1$, $E_{n}$ observes the environment to obtain $S_{n}(t+1)$. The quadruple $\left ( S_{n}(t), a_{n}(t), R_{n}(t), S_{n}(t+1) \right ) $ represents an interaction experience between the agent and the environment. It can be seen that $E_{n}$ continuously stores the experience until the memory buffer is full and then starts learning.

In the learning phase, $E_{n}$ takes $S_{n}(t) $ as input to the $Q_{eval}$ network to get $Q_{eval}\left ( S_{n}(t), A_{n}(t) \right )$ for all possible values of $A_{n}\left ( t \right )$. Taking $S_{n}(t+1)$ as the input to the $Q_{eval}$ network, we can get $Q_{eval}\left ( S_{n}(t+1), A_{n}(t) \right )$ for all possible values of $A_{n}\left ( t \right )$, and then choose the best value of $A_{n}\left ( t \right )$ as $A$, which makes the value of $Q_{eval}\left ( S_{n}(t+1), A \right )$ the largest. Taking $R_{n}(t)+\gamma\times Q_{target}\left ( S_{t+1}, A \right )$ as the real value of the network and $Q_{eval}\left ( S_{t}, A \right )$ as the predicted value of the network, we perform reverse error propagation, where $\gamma$ represents the discounted rate. We then reduce the value of $\epsilon$ to reduce the probability of random movements. After a few steps of learning, the parameters of the $Q_{eval}$ network are hard copied to the $Q_{target}$ net for parameter update.

\subsubsection{Tasks uploading}
After the server uses the reward evaluation algorithm to generate the scheduling result, the device performs local processing or offloading tasks based on the scheduling result. The bandwidth is shared equally between each device and the MEC server. The server sequentially adds the tasks received from the device to the end of the task queue, and then sequentially removes the tasks from the beginning of the task queue. After an offloaded task is completed, the server sends the processing result to the corresponding terminal device via a transmission link.

\begin{figure}[t] 
\centering
\captionsetup{font={footnotesize }}
\subfigure[]{
\includegraphics[width=0.74\linewidth]{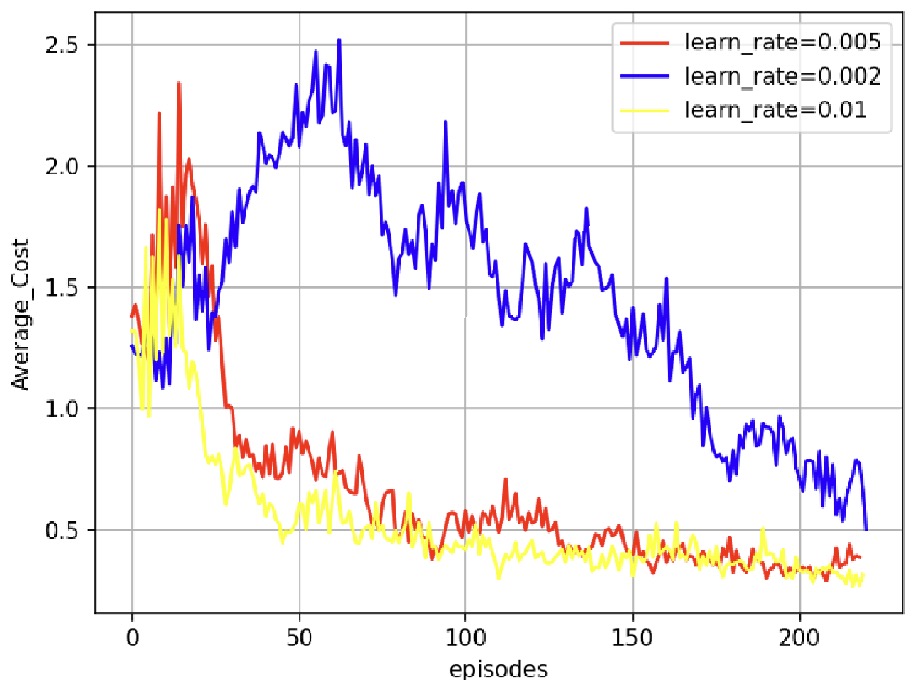}}
\hspace{-2mm}
\subfigure[]{
\includegraphics[width=0.74\linewidth]{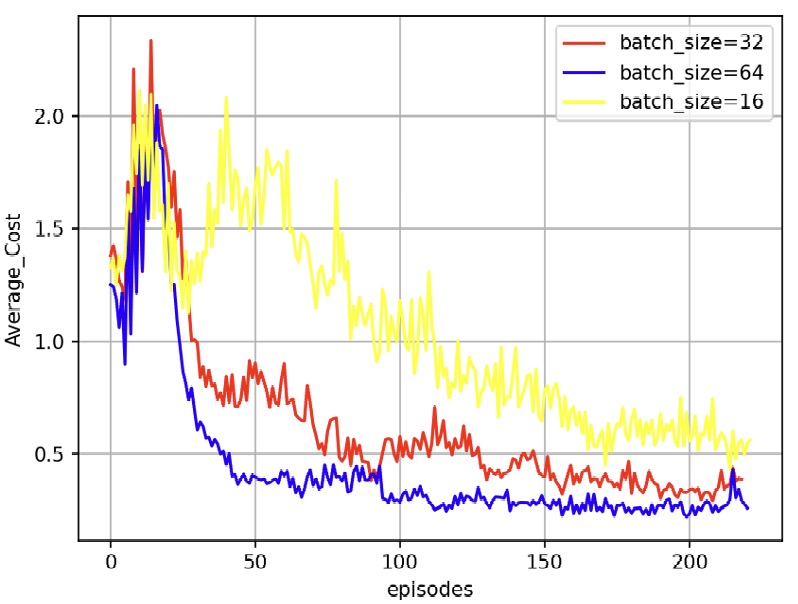}}
\caption{The average loss in different learning rates is shown in (a), the average loss in different batch sizes is shown in (b). }
\label{iterations} 
\end{figure}

\begin{figure}[t] 
\centering
\captionsetup{font={footnotesize }}
\subfigure[]{
\includegraphics[width=0.74\linewidth]{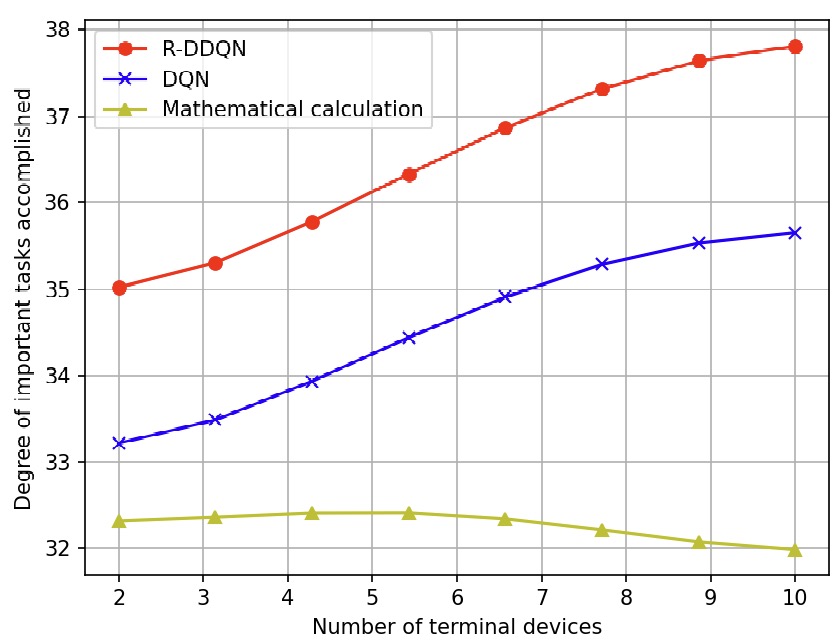}}
\hspace{-2mm}
\subfigure[]{
\includegraphics[width=0.74\linewidth]{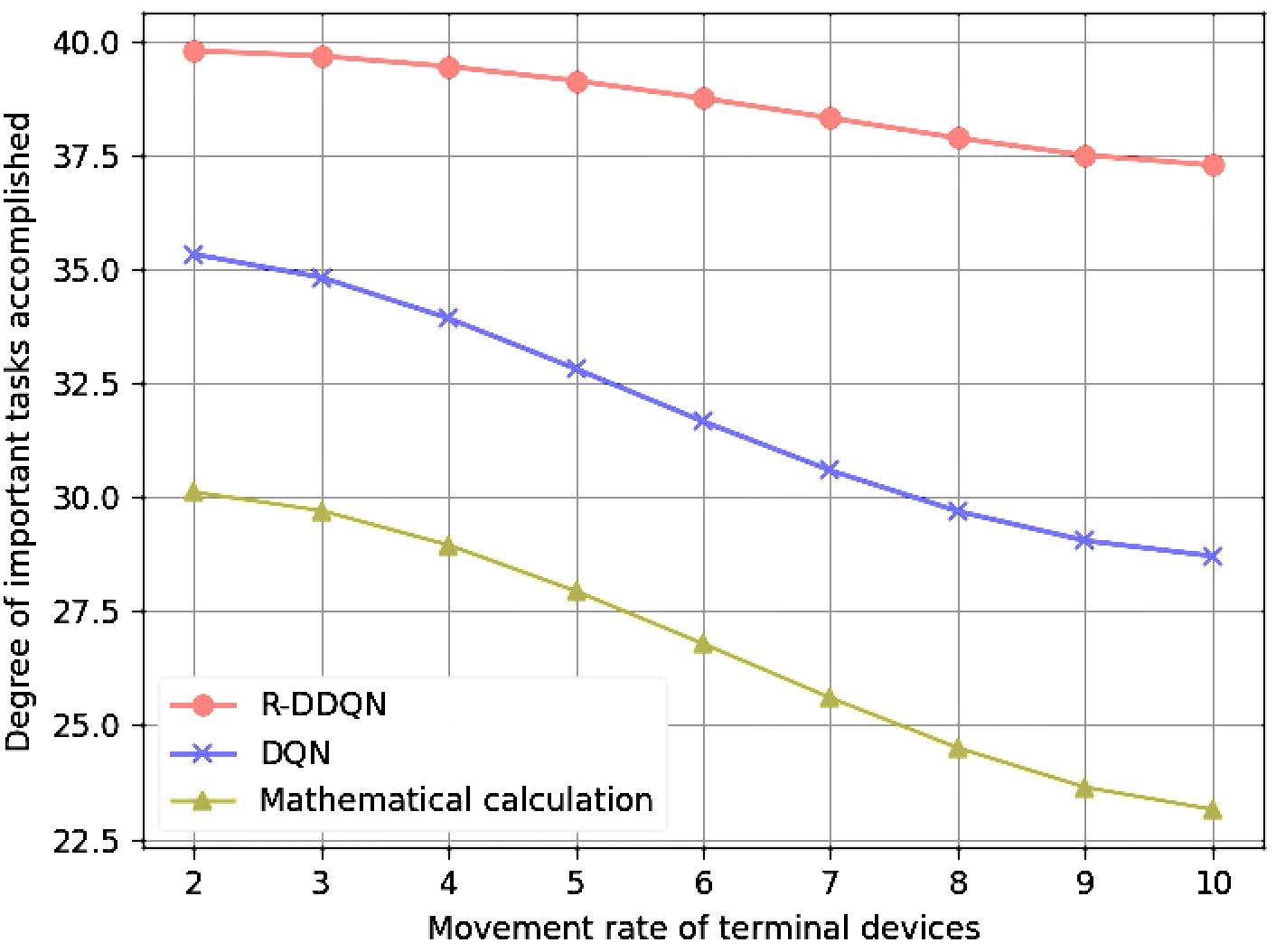}}
\caption{Performance evaluation of the proposed R-DDQN based algorithm versus different numbers of terminal devices and different movement rates of terminal devices is shown in (a) and (b), respectively. }
\label{performances} 
\end{figure}

\section{Experimental Results}

In the scenario considered, there are 3 MEC servers and 10 terminals. The terminal is set to move in a straight line, e.g. east, west, south and north. The mobility parameters change at the beginning of each time slot and the coordinates of the terminal are updated accordingly.

Fig.~\ref{iterations}(a) and Fig.~\ref{iterations}(b) show the convergence of the network under different learning rates and batch sizes. The x-axis is the number of training episodes. The Y-axis shows the average error between the predicted value and the true value. As shown in Fig.~\ref{iterations}(a), the network loss decreases fastest at a learning rate of 0.01. In the last 50 iterations, the networks with a learning rate of 0.01 tend to converge and the network loss has stabilised at a low value. As shown in Fig.~\ref{iterations}(b), the network loss decreases fastest and stabilises at a low value when the batch size is 64. Therefore, we set the learning rate and data batch size to 0.01 and 64, respectively, to ensure that the network converges with high speed and accuracy.

Fig.~\ref{performances}(a) and Fig.~\ref{performances}(b) show the comparison of our algorithm with DQN and the mathematical calculation method. The mathematical calculation method is to traverse the space of offloading strategies for revenue evaluation at a given depth and return the optimal strategy. The Y-axis represents the degree of completion of important tasks in unit time, denoted as $I$. In Fig.~\ref{performances}(a), the X-axis represents the number of terminals. As shown in the figure, comparing DQN and traditional mathematical algorithms, our proposed algorithm maintains good performance as the number of end devices increases. In Fig.~\ref{performances}(b), our algorithm has less performance degradation when the movement rate of the terminal device increases. Taken together, our algorithm is stable and does not lose much performance even when the number of terminal devices and the movement rate continue to increase.

\section{Conclusion}
This paper mainly studies the problem of computation offloading and request scheduling for non-divisible tasks in multi-server multi-access edge computing systems. We formulate an optimization problem and then design a reward evaluation algorithm based on DDQN to solve the offloading scheduling problem, which is solved by considering the characteristics of the task and the mobility of the terminals in the computational offloading domain. Finally, the numerical results show that the algorithm has an excellent performance, which can achieve the initial goal and more valuable tasks.

% {\appendix[Proof of the Zonklar Equations]
% Use $\backslash${\tt{appendix}} if you have a single appendix:
% Do not use $\backslash${\tt{section}} anymore after $\backslash${\tt{appendix}}, only $\backslash${\tt{section*}}.
% If you have multiple appendixes use $\backslash${\tt{appendices}} then use $\backslash${\tt{section}} to start each appendix.
% You must declare a $\backslash${\tt{section}} before using any $\backslash${\tt{subsection}} or using $\backslash${\tt{label}} ($\backslash${\tt{appendices}} by itself
%  starts a section numbered zero.)}
% \bibliographystyle{unsrt}
% \bibliography{ref.bib}


\end{document}